\documentclass[aps,pra, superscriptaddress,showpacs,showkeys]{revtex4}
\usepackage{amssymb,bm}
\usepackage{graphicx}
\usepackage{amsmath}
\allowdisplaybreaks

\begin{document}

\title{Charge asymmetry in  high-energy  $\mu^+\mu^-$ photoproduction  in the electric field of a heavy atom}

\author{E.J. Downie}
\address{The George Washington University, Washington, DC 20052, USA}
\author{R.N. Lee}
\author{A. I. Milstein}
\address{G.I.Budker Institute of Nuclear Physics, 630090 Novosibirsk, Russia}
\author{G. Ron}
\address{Hebrew University of Jerusalem, Jerusalem, 91904 Israel}

\date{\today}

\begin{abstract}
The charge asymmetry in the  differential cross section of  high-energy  $\mu^+\mu^-$ photoproduction  in the electric field of a heavy atom is obtained.
This asymmetry arises due to the Coulomb corrections to the amplitude of the process (next-to-leading term with respect to the atomic field). The deviation of
the nuclear electric field from the Coulomb field at small distances  is crucially important for the charge asymmetry. Though the Coulomb corrections to the total cross section
are negligibly small, the charge asymmetry is  measurable for selected final states of
  $\mu^+$ and $\mu^-$. We further discuss the
feasibility for experimental observation of this effect.
\end{abstract}



\maketitle

\section{Introduction}
Photoproduction of  muon pairs off heavy nuclei is one of the most interesting and important QED processes.
The Born approximation cross section is known for arbitrary energy $\omega$ of the incoming photon, Refs.~\cite{BH1934,Racah1934}
(we set $\hbar=c=1$ throughout the paper). The Born cross section is proportional to the square
of the nuclear form factor $F(Q^2)$ and is sensitive to its shape since, for  a heavy nucleus, the Compton wavelength of muon, $\lambda_\mu=1/m=1.87$~fm,  is less than the nuclear radius,
$R=7.3$~fm for gold and $R=7.2$~fm for lead, $m$ is the muon mass. Usually, for heavy atoms, one must account for the
higher-order terms in the perturbative expansion with respect to the parameter $\eta=Z\alpha$ (Coulomb corrections),
where $Z$ is the atomic charge number, $\alpha=e^2\approx 1/137$ is the fine-structure constant, and $e$ is the electron charge.
The Coulomb corrections to the total cross section of  muon pair photoproduction were discussed in a set of publications
\cite{IM98, HKS07, JS09}. In contrast to the Born cross section, where the main contribution is given by the impact parameter
$\rho$ in the region $R\ll\rho\ll \omega^2/m$, the main contribution to the Coulomb corrections stems from region
$\rho\sim \lambda_\mu\lesssim R$. Thus, the Coulomb corrections to the total cross section are strongly  suppressed by the form factor. Therefore,  one may expect
that this statement is also valid for all quantities related to the Coulomb corrections. In this paper we show that this is, in fact, not the case. We consider the charge
asymmetry in the differential cross section $d\sigma(\bm p,\,\bm q,\,\eta)$ of high energy $\mu^+\mu^-$ photoproduction off a heavy atom, where $\bm p$ and $\bm q$ are
the momenta of $\mu^-$ and $\mu^+$, respectively, $\omega\gg m$. The charge asymmetry $\cal A$ is defined as
\begin{eqnarray}\label{A}
&&{\cal A}=\frac{d\sigma_a(\bm p,\,\bm q,\,\eta)}{d\sigma_s(\bm p,\,\bm q,\,\eta)}\,,\,
d\sigma_s(\bm p,\,\bm q,\,\eta)= \frac{d\sigma(\bm p,\,\bm q,\,\eta)+d\sigma(\bm q,\,\bm p,\,\eta)}{2}\,,\,
d\sigma_a(\bm p,\,\bm q,\,\eta)= \frac{d\sigma(\bm p,\,\bm q,\,\eta)-d\sigma(\bm q,\,\bm p,\,\eta)}{2}\,.
\end{eqnarray}
 It follows from  charge parity conservation that $d\sigma(\bm p,\, \bm q,\, \eta)=d\sigma(\bm q,\, \bm p,\, -\eta)\,,$  so that
$ d\sigma_s(\bm p,\, \bm q,\, \eta)$ is an even function of $\eta$ and  $d\sigma_a(\bm p,\, \bm q,\, \eta)$ is an odd function of $\eta$.
For $\omega\gg m$, small angles between the vectors $\bm p$ , $\bm q$, and incoming photon momentum  $\bm k$, it is possible to make use of the quasiclassical approximation.
The Coulomb corrections for the high energy  $e^+e^-$ photoproduction cross section were obtained in the leading quasiclassical approximation in Refs.\cite{BM1954,DBM1954}.
In this case the the nuclear form factor correction is negligible, and for a heavy nucleus the terms to all orders in the parameter $\eta$ should be taken into account. However,  in the leading quasiclassical
approximation $d\sigma_a(\bm p,\,\bm q,\,\eta)=0$, i.e., charge asymmetry is absent. The Coulomb corrections to the spectrum and to the total cross section of $e^+e^-$ photoproduction in a strong atomic field
were derived in the next-to-leading quasiclassical approximation in Ref.~\cite{LMS2004}. The Coulomb corrections to the differential cross section were derived in the next-to-leading quasiclassical approximation in Ref.~\cite{LMS2012}, where the charge
asymmetry $\cal A$ was studied in detail in all orders in $\eta$. For $\mu^+\mu^-$ high energy photoproduction, the  structure of the Coulomb corrections to the differential cross section is different.
When the momentum transfer $Q_\perp\gtrsim 1/R$ the form factor dependence strongly suppresses the cross section. Here $\bm Q=\bm p+\bm q-\bm k$, $\bm Q_\perp=\bm Q-(\bm Q\cdot\bm \nu)\bm \nu$,
and $\bm \nu=\bm k/\omega$. Therefore,  to have the noticeable charge asymmetry and the noticeable cross section, we should consider the region  $Q_\perp\lesssim 1/R$, but $p_\perp\sim q_\perp\sim m\gg Q_\perp$, so that
$|\bm p_\perp+\bm q_\perp|\ll|\bm p_\perp-\bm q_\perp|$. As was shown in Ref.~\cite{LMS2012}, in this region only the term
$\propto\eta^3$ survives in the expansion of the Coulomb corrections in $\eta$ even for $\eta\sim 1$.
In the  present paper, we calculate $d\sigma_a(\bm p,\,\bm q,\,\eta)\propto\eta^3$ in the region $|\bm p_\perp+\bm q_\perp|\ll|\bm p_\perp-\bm q_\perp|$ taking into account the nuclear form factor correction.
This term gives rise to the charge asymmetry $\cal A\propto\eta$. We show that $\cal A$ and $d\sigma_s(\bm p,\,\bm q,\,\eta)\propto\eta^2$ are large enough to be observed
experimentally. The possibility of experimental observation of the charge asymmetry is discussed in detail. We also note that for  $p_\perp\gg m$ , and $q_\perp\gg m$,
$d\sigma_a$  was also investigated in Ref. \cite{BrGil68} in scalar electrodynamics.

\section{General discussion}

The cross section for $\mu^+\mu^-$ pair production by a high-energy photon in an
external field reads (see, e.g., Ref.~\cite{BLP82} )
\begin{equation}\label{eq:cs}
d\sigma= \frac{\alpha}{(2\pi)^4\omega}\,d\bm{p}_\perp\,d\bm{q}_\perp\,d\varepsilon_p\,|M_{\lambda_1\lambda_2\lambda_3}|^{2}
\,,
\end{equation}
where $\varepsilon_{ p}=\sqrt{ p^2+m^2}$, $\varepsilon_{ q}+\varepsilon_{ p}=\omega$,    $\bm p$ and $\bm q$ are the $\mu^-$ and $\mu^+$ momenta, respectively,
$\bm{p}_\perp$ and $\bm{q}_\perp$ are components of the vectors $\bm{p}$ and $\bm{q}$ perpendicular to the photon
momentum  $\bm k$  . The matrix element $M_{\lambda_1\lambda_2\lambda_3}$ has the form
\begin{equation}\label{M12}
M_{\lambda_1\lambda_2\lambda_3}\,=\,\int d\bm r \,\bar u_{\lambda_1\bm p }^{(out)}(\bm r )\,\bm\gamma\cdot
\bm e_{\lambda_3}\,v _{\lambda_2\bm q}^{(in)}(\bm r )\exp{(i\bm k\cdot\bm r )}\,\,.
\end{equation}
Here $ u_{\lambda_1\bm p}^{(out)}(\bm r )$ is a positive-energy  solution and $v_{\lambda_2\bm q}^{(in)}(\bm r )$ is a
negative-energy solution of the Dirac equation in the external field, $\lambda_1=\pm 1$ and $\lambda_2=\pm 1$ enumerate the independent solutions of the Dirac
equation, and $\lambda_3=\pm 1$ enumerates the photon polarization vector,
$\bm e_{\lambda_3}$,  $\gamma^\mu$ are the Dirac matrices.
Note that the asymptotic form of $ u_{\lambda\bm p}^{(out)}(\bm r )$
at large $\bm r$ contains the plane wave and the spherical convergent wave,
while the asymptotic form of  $ v_{\lambda\bm q}^{(in)}(\bm r )$ at large $\bm r$
contains the plane wave and the spherical divergent wave.

It is convenient to find the solutions of the Dirac equation in the atomic potential  $V(r)$, using the relations
(see , e.g., \cite{LMS2012})
\begin{eqnarray}\label{Green1}
&&\frac{\exp(ipr_2)}{4\pi r_2}\bar{u}_{\lambda{\bm p}}^{(out)}({\bm
r}_1)=-\lim_{ r_2\to \infty }\,\frac{1}{ 2 \varepsilon_{ p}}\bar{u}_{\lambda{\bm p}}\gamma^0
G({\bm r}_2,{\bm r}_1|\varepsilon_{ p}) \,,\quad \bm p=p\bm n_2\,
,\nonumber\\
&&\frac{\exp(ipr_1)}{4\pi r_1}v_{\lambda{\bm p}}^{(in)}({\bm
r}_2)=\lim_{ r_1\to \infty }\,\frac{1}{ 2 \varepsilon_{ p}}G({\bm r}_2,{\bm r}_1|-\varepsilon_{ p})
\gamma^0{v}_{\lambda{\bm p}} \,,\quad \bm p=p\bm n_1\,
,\nonumber\\
&& u_{\lambda{\bm p}}=\sqrt{\frac{\varepsilon_p+m}{2\varepsilon_p}}
\begin{pmatrix}
\phi_\lambda\\
\dfrac{{\boldsymbol \sigma}\cdot {\boldsymbol
p}}{\varepsilon_p+m}\phi_\lambda\end{pmatrix}\, ,\quad
v_{\lambda{\bm p}}=\sqrt{\frac{\varepsilon_p+m}{2\varepsilon_p}}
\begin{pmatrix}
\dfrac{{\boldsymbol \sigma}\cdot {\boldsymbol
p}}{\varepsilon_p+m}\chi_\lambda\\
\chi_\lambda\end{pmatrix}\, ,
\end{eqnarray}
where $\bm n_1=\bm r_1/r_1$,  $\bm n_2=\bm r_2/r_2$, and $G({\bm r}_2,{\bm r}_1|\varepsilon)$ is the Green function  of the Dirac equation in the atomic potential $V(r)$.
We express the wave functions via the asymptotics of the Green function $D(\bm r_2,\bm r_1 |\,\varepsilon )$
of the squared Dirac equation,
  \begin{equation}\label{D}
D(\bm r_2,\bm r_1 |\varepsilon )= \langle\bm r_2|\left[
(\varepsilon-V(r))^2- p^2-m^2+i\bm\alpha\cdot{\bm\nabla} V(r)+i0 \right]^{-1}|\bm r_1\rangle\,,
\end{equation}
where $\bm\alpha=\gamma^0\bm\gamma$.  Using the relation
\begin{eqnarray}\label{g2}
&&G(\bm r_2,\bm r_1 |\,\varepsilon )= \left[
\gamma^{0}(\varepsilon -V(r_2))\, +i\bm\gamma\cdot\underrightarrow{{\bm \nabla}}_2\, +\, m \right]
D(\bm r_2,\bm r_1 |\varepsilon )\, ,\nonumber\\
&&G(\bm r_2,\bm r_1 |\varepsilon )=D(\bm r_2,\bm r_1|\,\varepsilon ) \left[
\gamma^{0}(\varepsilon -V(r_1))\, -i\bm\gamma\cdot\underleftarrow{{\bm \nabla}}_1\, +\, m \right]
\, ,
\end{eqnarray}
where $\underrightarrow{{\bm \nabla}}_2$ denotes the gradient over $\bm r_2$ acting to the right, while
 $\underleftarrow{{\bm \nabla}}_1$ denotes the gradient over $\bm r_1$ acting to the left.
Using Eqs.(\ref{g2})  and (\ref{Green1}), we arrive at the following result for the wave  functions
\begin{eqnarray}\label{wfD}
&&\frac{\exp(ipr_2)}{4\pi r_2}\bar{u}_{\lambda{\bm p}}^{(out)}({\bm
r}_2)=-\lim_{ r_2\to \infty }\,\bar{u}_{\lambda{\bm p}}
D({\bm r}_2,{\bm r}_1|\varepsilon_{ p}) \,,\quad \bm p=p\bm n_2,\nonumber\\
&&\frac{\exp(ipr_1)}{4\pi r_1}v_{\lambda{\bm p}}^{(in)}({\bm r}_2)
=-\lim_{ r_1\to \infty }\,D({\bm r}_2,{\bm r}_1|-\varepsilon_{ p})
{v}_{\lambda{\bm p}} \,,\quad \bm p=p\bm n_1\,.
\end{eqnarray}
It follows from  Eqs. (\ref{D}) and (\ref{wfD})  that the wave functions
$\bar{u}_{\lambda_1{\bm p}}^{(out)}({\bm r})$ and  $v_{\lambda_2{\bm q}}^{(in)}({\bm r})$ have the form,
\begin{align}\label{wfD1}
&\bar{u}_{\lambda_1{\bm p}}^{(out)}({\bm r})=
\bar{u}_{\lambda_1{\bm p}}\left[f_0(\bm p,\bm r)-\bm\alpha\cdot\bm f_1(\bm p,\bm r)
-\bm\Sigma\cdot\bm f_2(\bm p,\bm r)\right]\,,\nonumber\\
&v_{\lambda_2{\bm p}}^{(in)}({\bm r})=\left[g_0(\bm q,\bm r)+\bm\alpha\cdot\bm g_1(\bm q,\bm r)
+\bm\Sigma\cdot\bm g_2(\bm q,\bm r)\right]v_{\lambda_2{\bm q}} \,.
\end{align}
The term with $\gamma^5$ does not appear because it is impossible to construct a pseudoscalar using two vectors,
$\bm p$ and $\bm r$. The functions $f_0(\bm p,\bm r)$, $\bm f_1(\bm p,\bm r)$, and $\bm f_2(\bm p,\bm r)$ may be
obtained from the corresponding functions
$g_0(\bm q,\bm r)$, $\bm g_1(\bm q,\bm r)$, and $\bm g_2(\bm q,\bm r)$ by the replacement $\bm q\rightarrow\bm p$ and $V(r)\rightarrow -V(r)$. Note that the perturbation expansion of the functions $f_0(\bm p,\bm r)$, $\bm f_1(\bm p,\bm r)$,
 and $\bm f_2(\bm p,\bm r)$  starts from the terms $V^0$, $V^1$ and $V^2$, respectively.

Let us introduce the quantities
\begin{eqnarray}\label{fg}
&&(A_{00},\,\bm A_{01},\,\bm A_{10},\,\bm A_{02},\,\bm A_{20})=\int \!\!d\bm r\,\exp{(i\bm k\cdot\bm r )}(f_0g_0,\, f_0\bm g_1,\, \bm f_1 g_0,\,f_0\bm g_2\,, \bm f_2 g_0)\,.
\end{eqnarray}
In terms of these quantities, we find
\begin{eqnarray}\label{M2}
S&=&\frac{1}{2}\sum_{\lambda_1,\lambda_2,\lambda_3=\pm 1}|M_{\lambda_1\lambda_2\lambda_3}|^{2}=2(S_0+S_1)\,,\nonumber\\
S_0&=&\frac{1}{4}\left[\left(\frac{m\omega}{\varepsilon_p\varepsilon_q}\right)^2
+\theta_p^2+\theta_q^2\right]|A_{00}|^2
+|\bm A_{01}|^2+|\bm A_{10}|^2+\mbox{Re}\,A_{00}^*\left(
\bm\theta_p\cdot\bm A_{10}+\bm\theta_q\cdot\bm A_{01}\right)\,,\nonumber\\
S_1&=&-\mbox{Im}\,\left\{
\left[\bm A_{20}^*\times\left(\bm\theta_p A_{00}+2\bm A_{10}\right)\right]\cdot\bm\nu
+\left[\bm A_{02}^*\times\left(\bm\theta_q A_{00}+2\bm A_{01}\right)\right]
\cdot\bm\nu\right\}\,,
\end{eqnarray}
where $\bm\nu=\bm k/\omega$, $\bm \theta_p=\bm p_{\perp}/\varepsilon_p$,  $\bm \theta_q=\bm q_{\perp}/\varepsilon_q$.
In deriving Eq.(\ref{M2}) we sum over the polarization of the
 $\mu^+$ and $\mu^-$  and average over the photon polarization. The expression for $S$ is very convenient for
further consideration. It is obtained in the  quasiclassical approximation  with the first quasiclassical correction taken into account. Both terms, the leading  term and the correction, are exact in the atomic field. For high energy $\mu^+\mu^-$ photoproduction, as it was discussed above,  for the symmetric part of the cross section it is sufficient to use the Born result  ($\propto V^2$), while for the antisymmetric part of the cross section we use the term $\propto V^3$. Note that the perturbation expansion of $A_{00}$, $\bm A_{10}$, and $\bm A_{01}$ starts from the terms $\propto V$, and the expansion of $\bm A_{20}$, and $\bm A_{02}$ starts from the terms $\propto V^2$.

\section{Calculation of the matrix elements and cross section}
\begin{figure}[htp]
\includegraphics[scale=0.6]{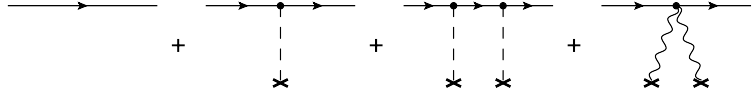}
\caption{Diagrams of the perturbation theory for the wave function. The dashed  line corresponds to the operator
 $2\varepsilon V(r)-i(\bm\alpha\cdot{\bm\nabla}) V(r)$, and seagull corresponds to the operator $-V^2(r)$.
}\label{dia}
\end{figure}

Using the conventional perturbation theory, see Eqs.~(\ref{D}), (\ref{wfD}), and Fig.\ref{dia},  we find for the terms linear in the potential,
\begin{eqnarray}\label{A1}
&&A_{00}^{(1)}= \frac{2\varepsilon_p\varepsilon_q V_F(Q^2)}{\omega m^2}(\xi_p-\xi_q)\,,\quad
\bm A_{01}^{(1)}=-\frac{\varepsilon_p V_F(Q^2)}{\omega m^2}\xi_p\bm Q\,,\quad
\bm A_{10}^{(1)}=\frac{\varepsilon_q V_F(Q^2)}{\omega m^2}\xi_q\bm Q\,,\nonumber\\
&&\xi_p=\frac{1}{1+\delta_p^2}\,,\quad \xi_q=\frac{1}{1+\delta_q^2}\,,\quad
\bm\delta_p=\frac{\varepsilon_p\bm\theta_p}{m}\,,\quad \bm\delta_q=\frac{\varepsilon_q\bm\theta_q}{m}\,,\quad
\bm Q=\bm p+\bm q-\bm k\,.
\end{eqnarray}
Here $V_F(Q)$ is the Fourier transformation of the potential $V(r)$, $V_F(Q^2)=-4\pi\eta{ F}(Q^2)/Q^2$, where
 ${F}(Q^2)$ is the form factor which differs essentially  from unity at $Q\gtrsim 1/R$ and $Q\lesssim 1/r_{scr}$,
 where $R$ is the nuclear radius and $r_{scr}$ is the screening radius. For $\mu^+\mu^-$ photoproduction,
 the effect of screening is negligible.

\begin{figure}[htp]
\includegraphics[scale=1.2]{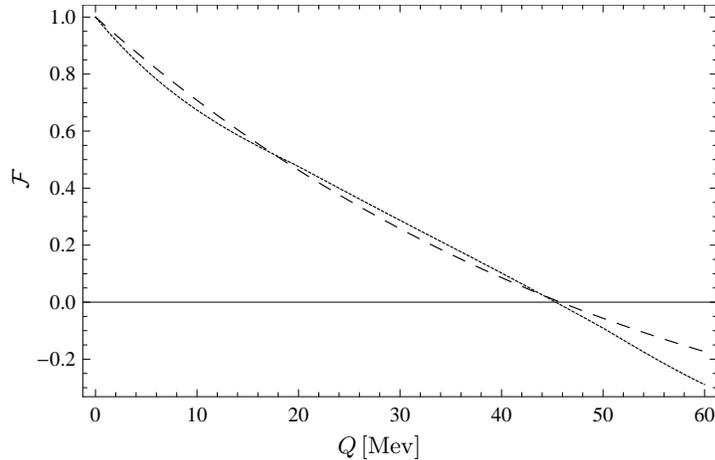}
\begin{picture}(0,0)(0,0)
\put(-150,-10){ $Q\,[\mbox{Mev}]$}
 \put(-270,90){\rotatebox{90}{${\cal F}$}}
 \end{picture}
\caption{The dependence of the function ${\cal F}(Q)$ on $Q$ for lead ( $Z=82$). The solid curve corresponds to the real charge distribution, the dashed curve is given by Eq.(\ref{calf}) with $\Lambda=60$~MeV.
}\label{figradius}
\end{figure}

From Eqs.(\ref{M2}) and (\ref{A1}) we find the well known result for  the leading term in $d\sigma_s$ (see, e.g.,\cite{BLP82}): \begin{eqnarray}\label{secsB}
&&d\sigma_s=\frac{2\alpha m^2\, d\varepsilon_p\,d\bm\delta_p\,d\bm\delta_q}{(2\pi)^4 \omega^3}{V_F}^2(Q^2)
\left[\frac{Q^2}{m^2}\xi_p\xi_q(\varepsilon_p^2+\varepsilon_q^2)+2\varepsilon_p\varepsilon_q(\xi_p-\xi_q)^2\right]\,.
\end{eqnarray}
We now calculate the next-to-leading quasiclassical correction to the cross section. This correction  is proportional to $V^3$ and arises  from the interference between the leading term of the matrix element $\propto V$ and
the next-to-leading term  $\propto V^2$. Since  $A_{00}^{(1)}$, $\bm A_{01}^{(1)}$, and $\bm A_{10}^{(1)}$ are the real quantities, one should calculate the real parts of  $A_{00}^{(2)}$, $\bm A_{01}^{(2)}$,  $\bm A_{10}^{(2)}$, and the imaginary parts of $\bm A_{02}^{(2)}$
and $\bm A_{20}^{(2)}$, see Eq. (\ref{M2}).  A straightforward calculation gives
\begin{eqnarray}\label{A2}
&&\mbox{Re}A_{00}^{(2)}= \frac{\varepsilon_p\xi_p+\varepsilon_q\xi_q}{\omega m^2}(J_0-J_1)\,,\nonumber\\
&&\mbox{Re}\bm A_{01}^{(2)}=\frac{\varepsilon_p\xi_p}{2\varepsilon_q\omega m^2}J_1\bm Q \,,\quad
\mbox{Re}\bm A_{10}^{(2)}=\frac{\varepsilon_q\xi_q}{2\varepsilon_p\omega m^2}J_1\bm Q \,,\nonumber\\
&&\mbox{Im}\bm A_{02}^{(2)}=\frac{\varepsilon_p\xi_p}{2\varepsilon_q\omega m^2}J_0[\bm\nu\times\bm Q]\,,\quad
\mbox{Im}\bm A_{20}^{(2)}=\frac{\varepsilon_q\xi_q}{2\varepsilon_p\omega m^2}J_0[\bm\nu\times\bm Q]\,,\nonumber\\
&&J_0=\int\frac{d\bm s}{(2\pi)^3}V_F(\chi_+)V_F(\chi_-)\,,\quad
J_1=\int\frac{d\bm s}{(2\pi)^3}(4s_\parallel^2-Q^2)V_F(\chi_+)V_F'(\chi_-)\,,\nonumber\\
&&\chi_\pm=(\bm s\pm \bm Q/2)^2\,,\quad s_\parallel=\bm s\cdot\bm Q/Q
\end{eqnarray}
where $V_F'(\chi)=\partial V_F(\chi)/\partial\chi$.
Using Eqs.(\ref{A1}), (\ref{A2}) and (\ref{M2}), we obtain the antisymmetric  part of the cross section,

\begin{eqnarray}\label{secaB}
d\sigma_a&=&\,\frac{\alpha m^2 d\varepsilon_p\,d\bm\delta_p\,d\bm\delta_q}{(2\pi)^4\omega^3}\,
\bigg\{(\xi_p-\xi_q)\bigg[4(\varepsilon_p\xi_p+\varepsilon_q\xi_q)+
\frac{\omega(\varepsilon_p^2+\varepsilon_q^2)}{\varepsilon_p\varepsilon_q}\bigg]\nonumber\\
&&+(\varepsilon_p-\varepsilon_q)\frac{(\varepsilon_p^2+\varepsilon_q^2)}{\varepsilon_p\varepsilon_q}\,\frac{Q^2}{m^2}
\xi_p\xi_q\bigg\}V_F(Q^2)(J_0-J_1)\,.
\end{eqnarray}
For the Coulomb field, the calculation yields $J_0=2\pi^2\eta^2/Q$, and $J_1=0$. Thus, our result is in agreement with the result obtained in Ref. \cite{LMS2012}. In the formula for the  charge asymmetry, ${\cal A}=d\sigma_a/d\sigma_s$, the dependence on the nuclear radius enters via the ratio $(J_0-J_1)/V_F(Q^2)$. Very often the form factor is approximated by the formula $F_0(Q^2)=1/(1+Q^2/\Lambda^2)$, where $\Lambda\approx 80$~MeV for heavy nuclei. This approximation gives an accurate result up to $60$~MeV.  In this case  the function
 ${\cal F}(Q)= -2(J_0-J_1)/(\pi\eta QV_F(Q^2))$ has the simple form
\begin{eqnarray}\label{calf}
{\cal F}(Q)&=&(1+x^2)\left[1+\frac{2}{\pi}\arcsin\left(\frac{x}{\sqrt{x^2+4}}\right)-
\frac{4}{\pi}\arcsin\left(\frac{x}{\sqrt{x^2+1}}\right)\right]-\frac{12x}{\pi (4+x^2)}\,,\quad x=\dfrac{Q}{\Lambda}\,.
\end{eqnarray}
At $Q\ll \Lambda$, we have ${\cal F}(Q)\approx 1-\displaystyle{\frac{6Q}{\pi\Lambda}}$, so that the function ${\cal F}(Q)$ diminishes rapidly
with increasing $Q$.  In Fig.\ref{figradius} we show the dependence of the function ${\cal F}(Q)$ on $Q$ for lead ($Z=82$). The solid curve corresponds to the real charge distribution, while the dashed curve is given by Eq.(\ref{calf}) with $\Lambda=60$~MeV.

For $Q\ll |\bm p_\perp-\bm q_\perp|$, the formula (\ref{secaB}) simplifies to,
\begin{eqnarray}\label{secaBas}
&&d\sigma_a=\,\frac{\alpha m^2 d\varepsilon_p\,d\bm\delta_p\,d\bm\delta_q}{(2\pi)^4\omega^2}\,
(\xi_p-\xi_q)\bigg[2(\xi_p+\xi_q)+
\frac{(\varepsilon_p^2+\varepsilon_q^2)}{\varepsilon_p\varepsilon_q}\bigg]
V_F(Q^2)(J_0-J_1)\,.
\end{eqnarray}
And we obtain for the charge asymmetry
\begin{eqnarray}\label{CAfinal}
&&{\cal A}=\,\frac{\pi\eta m\omega\kappa(\xi_p+\xi_q+B)}{4\varepsilon_p\varepsilon_q(B+\kappa^2\xi_p\xi_q)}{\cal F}(Q)\,,\nonumber\\
&& B=\frac{\varepsilon_p^2+\varepsilon_q^2}{2\varepsilon_p\varepsilon_q}\,,\quad
\kappa=\frac{m(\xi_q-\xi_p)}{Q\xi_p\xi_q}\,,
\end{eqnarray}
Let $\chi$ is the angle between the vectors $\bm p_\perp$ and $-\bm q_\perp$. In order to estimate $\cal A$, let us consider the region
of interest from the experimental point of view, $|\chi|\ll |\varepsilon_p-\varepsilon_q|/\omega\ll 1$ and
$|\theta_p-\theta_q|/|\theta_p+\theta_q|\ll |\varepsilon_p-\varepsilon_q|/\omega$. In this region,
\begin{eqnarray}\label{CAex}
&&{\cal A}=\,\frac{\pi\eta \theta(1+2\xi)}{(1+4\xi^2\delta^2)}
{\cal F}(\theta |\varepsilon_p-\varepsilon_q|)\,\mbox{sgn}(\varepsilon_p-\varepsilon_q) \,,\nonumber\\
&& \theta=\frac{1}{2}(\theta_p+\theta_q),\quad \delta=\frac{\omega\theta}{2m},\quad \xi=\frac{1}{1+\delta^2} \,,       \end{eqnarray}
and all of the dependence on $\varepsilon_p-\varepsilon_q$ is contained in the function $\cal F$. Since Eq.~(\ref{CAfinal})
is valid for all $\eta\lesssim 1$, the prefactor of  $\cal F$ in Eq.~(\ref{CAex})  can easily reach ten percent or more.

\section{Possibility of experimental observation}
\begin{figure}[ht]
\centerline{\includegraphics[width=0.55\textwidth]{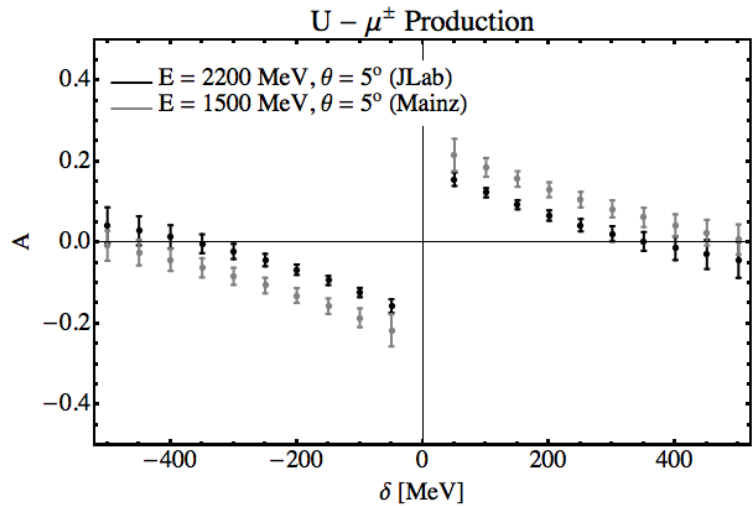}}
\caption{Calculated asymmetries and projected uncertainties for the experimental conditions
described in the text.}
\label{fig:exp}
\end{figure}

The above calculations clearly show that the size of the asymmetry is within reach of current experimental
capabilities, suggesting a possible measurement. Due to the low cross section, however, no current photon
facility has the required photon beam flux for such a measurement. We thus propose to make use of an electron
beam to provide the (virtual) photon flux, where we may calculate the equivalent photon flux close to  the
end-of-spectrum using the  approximation, see, e.g., Ref.~\cite{BFKK81}:
\begin{equation}
\label{eq:quasireal}
N_\gamma=\frac{\alpha}{\pi}\frac{\Delta}{E}\ln\left(\frac{\Delta}{m_e}\right)N_e,
\end{equation}
where $N_\gamma$($N_e$) is the photon (electron) flux, $E$ is the electron
beam energy, $m_e$ is the electron mass, and $\Delta$ is the region of integration around the endpoint. Eq.~(\ref{eq:quasireal})
is valid for $m_e\ll\Delta\ll E$. We identify two facilities with experimental capabilities suitable for the
proposed measurement. Those are the experimental Hall A at the Thomas Jefferson National
Accelerator Facility~\cite{Alcorn:2004sb}, and the A1 experimental hall at the Mainzer Mikrotron~\cite{Blomqvist:1998xn}.
We work in a region where the muon angles are  equal, $\theta_\mu^+=\theta_\mu^-$, and the
the sum of the muon energies is close to the beam energy, $E_\mu^+ + E_\mu^- \sim E$, so that the momentum
transfer to the recoiling nucleus is minimal, making the asymmetry essentially independent of the nuclear
form factor. The proposed kinematic conditions for both facilities are summarized in Table~\ref{tab:kinematics}.
\begin{table}[htb]
\begin{center}
\begin{tabular}{|c|c|c|} \hline
& JLab& Mainz \\ \hline
Beam Energy& 2.2 GeV& 1.5 GeV\\
Current& 50 $\mu$A & 50 $\mu$A\\
Detector Package& HRS + Septa magnets (see text)& Dedicated (see text)\\
Detector Angle& $5^\circ$& $5^\circ$\\
Target& $^{238}$U (25 $\mu$m)& $^{238}$U (25 $\mu$m)\\
\hline
\end{tabular}
\end{center}
\caption{\label{tab:kinematics}Proposed kinematical conditions.}
\end{table}
We assume a conservative solid angle of 0.3 msr for each of the detectors, an energy bin of 10 MeV, and
select events where the sum of the muon energies is within 10 MeV of the beam energy,
\begin{equation}
E-(E_{\mu^+}+E_{\mu^-})\le 10 {\rm MeV}.
\end{equation}
Figure~\ref{fig:exp} shows the calculated asymmetry and projected asymmetry as a function of $\delta=E_{\mu^+}-E_{\mu^-}$
for the aforementioned kinematics, where we assume 3h of beam time for each of the data points.
The proposed JLab experimental setup is essentially identical (except for the target) to the already
approved JLab experiment E12-10-009 (APEX)~\cite{APEX}, searching for
massive vector bosons (dark photons)~\cite{Bjorken:2009mm}. Thus, the proposed measurement can be trivially
conducted jointly with the APEX experiment. The MAMI/A1 detector setup is currently unsuitable
for the proposed experiment, due to the constraints on the possible detector angles, thus, a dedicated
detector setup would be required. Due to the relaxed requirement on the particle detection (muons
with energies between about 500 MeV and 1 GeV and a small solid angle) such a detector
setup is relatively easy to construct or obtain, as an example we mention the di-electron
production experiment, currently scheduled at the HIGS facility, which makes use of an appropriate
detector setup and which is expected to conclude data taking during 2014 or 2015.
Fig. \ref{fig:exp} clearly demonstrates the viability of such an experiment, which will be the first
to accurately measure di-muon production off heavy nuclei, where the parameter $\eta$ is not small. Also note
that in these experimental conditions it is possible to observe a second sign reversal of the asymmetry,
which happens due to cancellation in the function ${\cal F}(Q)$ (see Fig. \ref{figradius})

\section{Conclusion}
We have derived  the charge asymmetry ${\cal A}$ in the process  of $\mu^+\mu^-$ photoproduction in the electric field of
a heavy atom. This asymmetry is related to the first quasiclassical correction to the differential cross section of the process.
In the experimental region of interest, where $Q\ll |\bm p_\perp-\bm q_\perp|$, $Q\sim 1/R$, and
$ p_\perp\sim q_\perp\sim 1/m$, the asymmetry ${\cal A}$ can be as large as  a few tens of percent.
In this region our result is valid even for $\eta\sim 1$. Since $\lambda_\mu\lesssim R$, the charge asymmetry is
very sensitive to the shape of the nuclear form factor
and may be used to validate or perform measurements of these form factors. Additionally,
measurements of ${\cal A}$ can be used to investigate not only the nuclear form factor,
but also to search for new massive particles such as dark photons~\cite{Bjorken:2009mm}, and by
comparing results from electron and muon production, test lepton universality.
Finally, we have demonstrated that the experimental observation of the charge asymmetry in $\mu^+\mu^-$ photoproduction
in the electric field of heavy atoms is a realistic task
and suggested an experimental configuration which will allow such a measurement.

\acknowledgments
The work of R.N.L. and A.I.M. has been supported in part  by the Ministry of Education
and Science of the Russian Federation. This work has also been funded, in part, by the US National
Science Foundation (Grant No. 1309130). A.I.M.  thanks the Lady Davis Fellowship Trust and the Racah Institute of Physics at the Hebrew University  of Jerusalem for financial support and kind hospitality.

\end{document}